\documentclass[12pt,epsfig]{article}
\usepackage{epsfig}
\newcommand{\singlespace}{
    \renewcommand{\baselinestretch}{1}\large\normalsize}

\begin{document}

\begin{titlepage}
\pagestyle{empty}
\setcounter{page}{1}
\vspace{1.0in}
\begin{center}
\singlespace
\begin{large}
{\bf Kaons and pions  in strange quark matter}\\
\end{large}
\end{center}
\vskip 1.0in
\begin{center}
{\large {\bf Pedro Costa}\footnote{E-mail: pcosta@fteor5.fis.uc.pt} and 
{\bf Maria C. Ruivo}\footnote{E-mail: maria@teor.fis.uc.pt}}\\
\vskip0.5cm
Centro de F\'{\i}sica Te\' orica, Departamento de F\'{\i}sica\\
Universidade de Coimbra \\P-3004-516 Coimbra, Portugal\\ 
\end{center}
\vspace{2cm}

\begin{abstract}
\noindent
The behavior of kaons and pions in  strange quark matter in weak equilibrium, is investigated within the SU(3) Nambu-Jona-Lasinio [NJL] model. This work focuses a region of high densities where the behavior of mesons has not been explored in the framework of this model. It is found that above  the density where strange valence quarks appear in the medium, $\rho\,=\,3.8 \rho_0$, a change of behavior of different observables is observed indicating a tendency to the restoration of flavor symmetry.  In  particular, the 
splitting between charge multiplets,   $K^+,\,K^-;\,K^0,\,{\bar K^0}\,\mbox{and}\,\, \pi^+,\,\pi^-$ decrease and the  low energy modes with quantum numbers of $K^-,\,\bar K^0\,\mbox{and}\,\, \pi^+$, which are excitations of the Fermi sea,  are less relevant than for lower densities. 
\bigskip

\noindent
PACS: 11.30.Rd; 11.Hx; 14.40.Aq\\
Keywords: Strange quark matter; Kaons sum rules; Kaons masses; Pions masses

\end{abstract}
\end{titlepage}
\newpage

\section{Introduction}

It is generally accepted that  under extreme conditions (high density and /or temperature) hadronic matter undergoes two phase transitions, characterized respectively by deconfinement and  restoration of chiral symmetry. The  QCD vacuum would then be described by a weakly interacting gas of massless quarks and gluons. Major theoretical and experimental efforts  have been dedicated to heavy-ion physics looking for signatures of the quark gluon plasma \cite{kanaya,rhic,cern}. Special attention has also been given to neutron stars, which provide a natural  laboratory to study hadronic matter at high densities.

Theoretical concepts  on the behavior of matter at high densities have been mainly developed from  model calculations. The
Nambu-Jona-Lasinio [NJL]\cite{NJL}  type models have been extensively used
over the past years to describe low energy features of hadrons and also to
investigate restoration of chiral symmetry with temperature or density \cite
{Hatsuda94,RuivoSousa,SousaRuivo,Hiller,RSP,Ruivo,buballa}. 

Studies of   flavor asymmetric matter at high densities are also motivated  
 by the interest  in investigating  the behavior of mesons. As a matter of fact \cite{Kaplan}
 the charge multiplets of mesons, that are degenerated in
vacuum or in symmetric matter ($\rho_u\,=\,\rho_d\,=\rho_s$) are expected to have a splitting in flavor
asymmetric matter. In particular,  the masses of kaons (antikaons) would  increase (decrease) with density and  a similar effect would occur for $\pi^-$ and $\pi^+$ in neutron matter. 

A slight raising of the $K^+$ mass and a  lowering of the $%
K^-$ mass \cite{RSP,waas,Cassing,Schaffner} seems to be compatible  with  the analysis of data on kaonic atoms \cite{Friedmann94} and with the results of KaoS and FOPI collaborations at GSI \cite
{Herrman,Barth}. From the theoretical point of view, the driving mechanism for the mass
splitting is attributed mainly to the selective effects of the Pauli
principle, although, in the case of $K^-$, the interaction with the $\Lambda
(1405)$ resonance plays a significant role as well, at least as low densities are concerned \cite{waas}. 

A model aiming at describing hadronic behavior in the medium should account for the great variety particle-hole excitations that the medium can exhibit, some of them with the same quantum numbers as the hadrons under study \cite{kubodera}.  It has been established, within the framework of NJL models, the presence, in flavor asymmetric media, of low energy pseudoscalar modes, which are excitations of the Fermi sea \cite{RuivoSousa,SousaRuivo,Hiller}.  Such studies were  carried out   in quark matter simulating nuclear matter $(\rho_u = \rho_d\,, \rho_s=0)$, for charged kaons,  and neutron matter without beta equilibrium, for charged pions. The role of the 't Hooft interaction was not taken into account in the last case. The combined effect of density and temperature, as well as the effect of vector interaction, was discussed for the case of charged kaons. \cite{RSP,Ruivo,costa}. Only densities below $\sim 3 \rho_0$ were considered. 

The aim of this work is  to explore  the high density  region ($\rho > \rho_c=2.25 \rho_0$) of  quark matter  simulating neutron matter in weak equilibrium,  having in mind  the study of  the behavior of kaonic (charged and neutral) and pionic (charged) excitations. The work is carried out in the framework of the  SU(3) NJL model. Since for high densities the quarks are supposed to be deconfined  the lack of confinement of NJL model is not a drawback in this region. 

%%%%%%%%%%%%%%%%%%%%%%%%%%%%%%%%%%%%%%%%%%%%%%%%%%%%%%%%%%%%%%%%%%%%%%%%%%%%%%%%%%%%%%%%%%%%%%%%%%%%%%%%%%%%%%%%%%%%%%%%

\section{Formalism}

We work in a flavor SU(3) NJL type model with scalar-pseudoscalar
pieces and a determinantal term, the 't Hooft
interaction, which breaks the $U_A(1)$ symmetry. We use the following
Lagrangian: 
\begin{equation}
\begin{array}{rcl}
{\cal L\,} & = & \bar q\,(\,i\, {\gamma}^{\mu}\,\partial_\mu\,-\,\hat m)\, q+%
\frac{1}{2}\,g_S\,\,\sum_{a=0}^8\, [\,{(\,\bar q\,\lambda^a\, q\,)}%
^2\,\,+\,\,{(\,\bar q \,i\,\gamma_5\,\lambda^a\, q\,)}^2\,] \\[4pt] 
& + & g_D\,\, \{\mbox{det}\,[\bar q\,(\,1\,+\,\gamma_5\,)\,q\,] + \mbox{det}%
\,[\bar q\,(\,1\,-\,\gamma_5\,)\,q \,]\, \} \label{1} \\ 
&  & 
\end{array}
\label{eq:lag}
\end{equation}
%%%
The model parameters, the bare quark masses $m_d=m_u, m_s$, 
the coupling constants and the cutoff in three-momentum space, $\Lambda$, 
are  fitted to the experimental
values of  the masses of pseudoscalar mesons and $f_\pi$. Here we use the same parameterization 
as  \cite{RKH}, $\Lambda=602.3$ MeV, $g_S\,\Lambda^2=3.67$%
, $g_D\Lambda^5=-12.39$, $m_u=m_d=5.5$ MeV and $m_s=140.7$ MeV.

The six quark interaction can be put in a form suitable to use the
bosonization procedure (see \cite{Vogl,Ripka,Ruivo}):

\begin{equation}
{\cal L_D}\,=\, \frac{1}{6} g_D\,\, D_{abc} \,(\bar q\, {\lambda}%
^c\,q\,)\,[\,(\,\bar q\,\lambda^a\, q\,)(\bar q\,\lambda^b\, q\,) -
3\,(\,\bar q \,i\,\gamma_5\,\lambda^a\, q\,)\,(\,\bar q
\,i\,\gamma_5\,\lambda^b\, q\,)\,]
\end{equation}

\noindent with: $D_{abc}=d_{abc}\,, a,b,c\,\, \epsilon\, \{1,2,..8\}\,, %
\mbox{(structure constants of SU(3))}\,,D_{000}=\sqrt{\frac{2}{3}}\,,
D_{0ab}=-\sqrt{\frac{1}{6}}\delta_{ab}$.

The usual procedure to obtain a four quark effective interaction from this
six quark interaction is to contract one bilinear $(\bar q\,\lambda_a\,q)$.
Then, from the two previous equations, an effective Lagrangian is obtained:

\begin{eqnarray}
L_{eff}\,&=& \bar q\,(\,i\, {\gamma}^{\mu}\,\partial_\mu\,-\,\hat m)\, q \,\,
\nonumber \\
&+& S_{ab}[\,(\,\bar q\,\lambda^a\, q\,)(\bar q\,\lambda^b\, q\,)]
+\,P_{ab}[(\,\bar q \,i\,\gamma_5\,\lambda^a\, q\,)\,(\,\bar q
\,i\,\gamma_5\,\lambda^b\, q\,)\,],
\end{eqnarray}

\noindent where: 
\begin{eqnarray}
S_{ab}\,=\,g_S\,\delta_{ab}\,+\,g_D D_{abc} \,< \bar q\, {\lambda}^c\,q\,>,
\nonumber \\
P_{ab}\,=\,g_S\,\delta_{ab}\,-\,g_D D_{abc} \,< \bar q\, {\lambda}^c\,q\,>.
\end{eqnarray}

By using the usual methods of bosonization one gets the following effective
action: 
\begin{eqnarray}  \label{action}
I_{eff}=&-i&Tr\ {\rm ln}(\,i\,\partial_\mu \gamma_{\mu}-\hat
m+\sigma_a\,\lambda^a +i\,\gamma_5\, {\phi}_a\,\,{\lambda}^a\,)  \nonumber \\
&-&\frac{1}{2}(\,\sigma_a\,{S_{ab}}^{-1}\,\sigma_b\,+{\phi}_a\, {\ P_{ab}}%
^{-1}\,\phi_b\,).
\end{eqnarray}

In order to introduce medium effects, we use the thermal Green functions in the standard way \cite{RKH} and restrict ourselves to $T=0$. The following gap equations are obtained:

\begin{equation}
M_i\,=\,m_i\,-2\,g_S\,<\bar q_i\, q_i>\,-\,2\,g_D\,<\bar q_j\, q_j><\bar
q_k\, q_k>
\end{equation}

\noindent with $i\,,j\,,k$ cyclic and $<\bar q_i\, q_i>\,=\, - 4 M_i I_{1,i}$ are the quark condensates.

The poles in the kaon propagator are obtained from the condition $(\,1\,-\,K_P\,\,J_{PP}\,)\,=\,0$ with:

\begin{equation}
J_{PP}=4(I_{1,u}+I_{1,s}+[\omega^{2}-(M_{u}-M_{s})^{2}]I_{2}(\omega))\,,
\end{equation}
where $K_P\,=\,g_S\,+\,g_D \,<\bar d d>$,  
$\omega= q_0 \pm (\mu_s - \mu_u)$ is the excitation energy, which, for kaons at rest is equal to $ \pm m_{K^{\pm}}$  and the integrals are given by:

\begin{equation}
I_{1,i}=\frac{N_{c}}{4\pi ^{2}}\int_{\lambda _{i}}^{{\large \Lambda }}%
\frac{k^{2}}{E_{i}}dk,  
\end{equation}   
where $\lambda _{i}$ is the Fermi momentum, and
\begin{eqnarray}\label{I_2}
I_{2}(\omega) &=&-\frac{N_{c}}{4\pi ^{2}}\left\{  \int_{\lambda
_{u}}^{\Lambda }\frac{k^{2}}{[(E_{u}+\omega)^{2}-E_{s}^{2}]E_{u}}dk
+\int_{\lambda _{s}}^{\Lambda }\frac{k^{2}}{%
[(E_{s}-\omega)^{2}-E_{u}^{2}]E_{s}}dk\right\}. 
\end{eqnarray}
 Similar expressions are obtained for $K^0,\,\bar K^0$, by replacing  $u\leftrightarrow d$ and for $\pi^+\,,\pi^-$, by replacing $%
s\leftrightarrow d$.

%%%%%%%%%%%%%%%%%%%%%%%%%%%%%%%%%%%%%%%%%%%%%%%%%%%%%%%%%%%%%%%%%%%%%%%%%%%%%%%%%%%%%%%%%%%%%%%%%%%%%%%%%%%%%%%%%%%%%%%%

\section{Discussion and conclusions} 
At variance with \cite{RuivoSousa,SousaRuivo,Hiller,RSP,Ruivo,costa}, we  consider here the case of asymmetric quark matter with strange quarks in chemical 
equilibrium maintained by weak interactions and with charge neutrality,  by imposing the following constraints on the chemical potentials of quarks and electrons and on its  densities: 

\begin{equation}
\mu_{d}=\mu_{s}=\mu_{u}+\mu_{e},\,\, \mbox{ and }\,\,\,\frac{2}{3}\rho_{u}-\frac{1}{3}(\rho_{d}+\rho_{s})-\rho_{e}=0,
\end{equation}
with 
\begin{equation}
\rho_{i}=\frac{1}{\pi^{2}}(\mu_{i}^{2}-M_{i}^{2})^{3/2}\theta(%
\mu_{i}^{2}-M_{i}^{2}),\,\,\mbox{ and }\,\,\,\rho_{e}=\frac{\mu_{e}^{3}}{3\pi^{2}}.
\end{equation}

As discussed by several authors, this  version of the NJL model exhibits a first order phase transition \cite{RSP,buballa,costa}. As shown in \cite{buballa}, by using a convenient parameterization  the model may be interpreted as having a mixing phase --- droplets of light $u\,,d$ quarks with a density $\rho_c=2.25 \rho_0$ surrounded by a non-trivial vacuum --- and, above the 
critical density, a quark phase with partially restored chiral symmetry. As a matter of fact, the energy per particle of the quark system has two minimum, corresponding to the zeros of the pressure (see Fig. 1-a)), the minimum at  $\rho_c=2.25 \rho_0$  being an absolute minimum.

The dynamical quark masses and chemical potentials  are plotted in Fig. 1-b) as functions of $\rho_B$. As shown in \cite{buballa}, chiral symmetry is partially restored in the SU(2) sector, the $u\,\mbox{and}\, d$ quarks masses decreasing sharply.  An important point to be noticed is that, due to the 't Hooft contribution in the gap equations, the mass of the strange quark decreases smoothly and becomes lower than the chemical potential  at densities above 3.8$\rho_0$, which we will denote from now on as $\rho'_s$. A more pronounced decrease is then observed, which is due to the presence of $s$ quarks in this regime. As it will be discussed below, this fact is related to a change in the behavior of different observables, as compared to the region of lower densities. At $\rho =\rho'_s$ (Fig. 1-b)) the difference between the  chemical potentials $\mu_d=\mu_s$ has a maximum and, as the density increases there is 
a  tendency to the restoration of flavor symmetry that can  be seen as well in quark densities ( Fig. 1-c)) and mesonic masses (Fig. 2).  

Let us analyze the results  for the masses of $K^+,\,K^-,\, K^0,\,\bar K^0$ and for $\pi^+,\,\pi^-$ (Fig. 2 a)-c)).
In order to understand these results, it is useful to look for the limits of the Dirac and of the Fermi sea continua of $\bar qq$ excitations with the quantum numbers of the mesons under study. They can be obtained by inspection of the meson dispersion relations  (looking for the limits of the regions  of poles in  the integrals $I_2 (\omega)$, Eq. \ref{I_2}) and are plotted in Fig. 2 a) and b)  (dashed point lines): $\omega'= \sqrt{M^2_s + \lambda^2_s}+\sqrt{M^2_{u(d)}+ \lambda^2_s}$ is the lower limit of the Dirac continuum,  and $\omega_{up}= \sqrt{M^2_s + \lambda^2_s} - \sqrt{{M^2}_{u(d)} + {\lambda^2}_s}$, $\omega_{low}= \sqrt{{M^2}_s+ {\lambda^2}_{u(d)}}-\mu_{u(d)}$ are the upper and lower limits of the Fermi sea continuum.  We do not show the corresponding limits for the pions because, in the range of densities studied, the pion modes  remain outside   the continuum.

As usual, two kinds of solutions for mesons masses may be  found, corresponding to excitations of the Dirac sea modified by the presence of the medium, and to excitations of the Fermi sea, respectively; however there are significant differences in their  behavior at low and high densities.  Concerning the Dirac sea excitations, we observe the expected  splitting between 
 multiplets: the increase of the masses of $K^+,\,K^0\,\mbox{and}\,\,\pi^- $ with respect to
those of $K^-,\,\bar K^0\,\mbox{and}\,\,\pi^+$ is due to Fermi blocking and is more pronounced for kaons than for pions because there are $u$ and $d$ quarks in the Fermi sea and therefore there are repulsive effects due to the Pauli
principle acting on $\pi^+$.  One notices also that at the critical density,   $2.25\rho_0$ the antikaons enter in the Dirac continuum and became resonances, a behavior already observed in \cite{RSP}. The new findings are that at $\rho=\rho'_s$ there is a sharp increase of $\omega'$ and decrease of $\omega_{up}$ and the antikaons turn again in bound states and, while at low densities  the splitting increases with density,  at high densities it decreases  and the modes show a tendency to become degenerated again.

Below the lower limit of the Fermi sea continuum there are low bound states, denoted by the subscript $S$, with quantum numbers of $K^-,\,\bar K^0,\,\pi^+$, respectively, which are particle-hole excitations of the Fermi sea. In the last case the low energy mode   only exists  up to $\rho=\rho_c$. 

Analysis of the energy weighted sum rules allows to calculate the strength located in each mode \cite{RSP}.  The sum of the contributions of the three discrete solutions (dashed curves in Fig. 3) deviates from 1 as the density increases, indicating that the modes of the continuum became more relevant with increasing density, a behavior already found in \cite{RSP}.  We can also see from Fig. 3 that   above $\rho'_s$  the lower energy modes are less coupled to the quarks and its strength decrease, which means that they become less relevant  and that the opposite occurs for the antikaons that excitations of the Dirac sea. It is a well known feature in many-body physics that the Fermi sea continuum of excitations induces attractive correlations below its lower limit and repulsive above the upper limit \cite{Kaplan}. As seen in Fig. 2- a)-b), this continuum becomes very narrow above $\rho'_s$,  which explains the behavior mentioned above. 

Finally, we remark that those low energy modes only appear associated to a first order phase transition, as  it was shown in \cite{RSP,costa}. We do not observe kaon condensation, probably due to the simplicity of the model. In fact, the model, in its present form, does not allow to consider color superconductivity, which  is though to play a significant role on kaons. On the other side, our treatment  is the leading order calculation in the inverse number of colors, $1/N_C$,  which means that we use the Hartree approximation to calculate the energy of the quark system and RPA to calculate the meson masses. As pointed out by several authors,  the quantum fluctuations due to meson-like excitations are expected to be relevant for calculation of the quark masses and  certainly will have effects in the EOS. This study can be an useful starting point to investigations along theses lines.
 
\begin{center}
{\large Acknowledgement:}
\end{center}
Work supported by FCT, by project PRAXIS/P/FIS/12247/98, by grant SFRH/BD/3296/2000 (Pedro Costa) and GTAE.

%%%%%%%%%%%%%%%%%%%%%%

%%%%%%%%%%%%%%%%%%%%%%%%%%%%%%%%%%%%
\begin{figure}[h]
\begin{center}
\epsfig{file=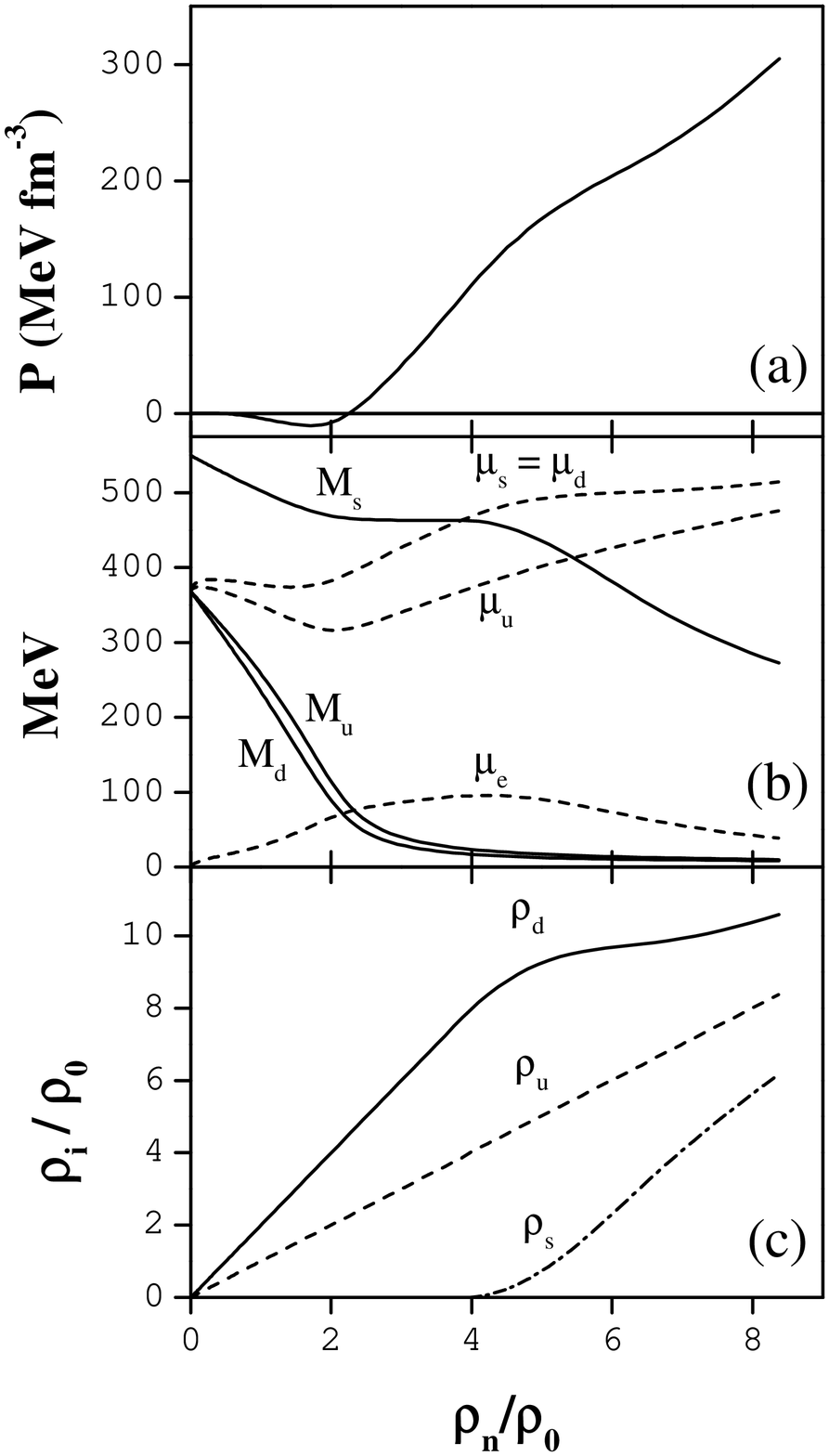,width=16.0cm,height=19.0cm}
\caption{Behavior of properties of quark matter in weak equilibrium as functions of the baryonic density. a) The pressure has two zeros, respectively for $\rho=0\, \mbox{and} \,\rho= 2.25 \rho_0$. b) Constituent quark masses (solid lines) and chemical potentials, $\mu_s=\mu_u\,, \mu_e$ (dashed lines). c) Densities of constituent quarks. $\rho_s \neq 0$ only above $4 \rho_0$, when $M_s < \mu_s$.}
\end{center}
\end{figure}
%%%%%%%%%%%%%%%%%%%%%%%%%%%%%%%%%%%%

%%%%%%%%%%%%%%%%%%%%%%%%%%%%%%%%%%%%
\begin{figure}[h]
\begin{center}
\epsfig{file=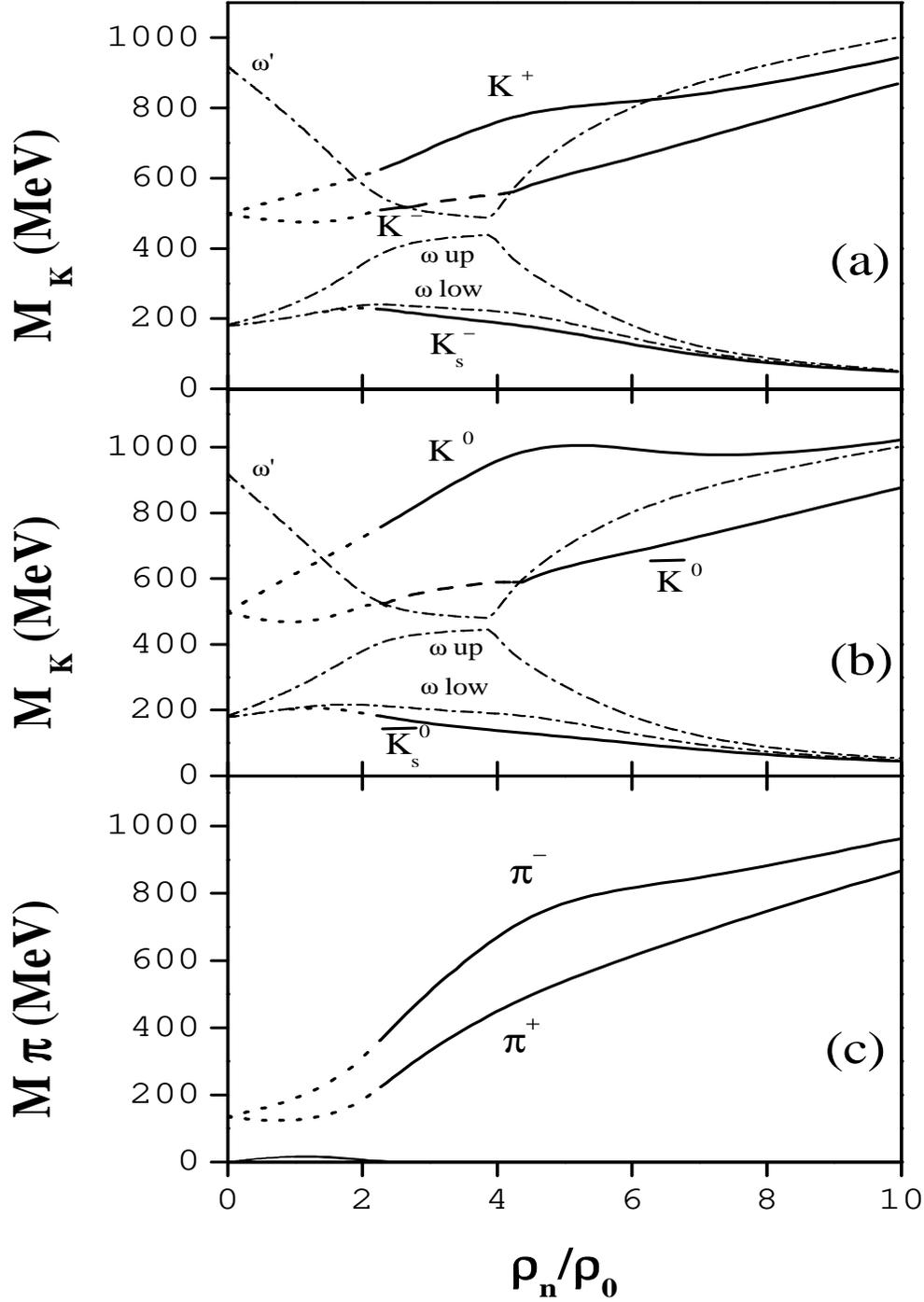,width=16.0cm,height=19.0cm}
\caption{Mesonic masses as functions of the baryonic density. a) Masses of $K^+,\,K^-$ and of the particle-hole excitation of the Fermi sea, $K^-_s$. Pointed lines represent values for $\rho < \rho_c=2.25 \rho_0$, which have no physical meaning, 
and dashed lines are for values in the  continuum. Dashed pointed lines represent  the lower limit of the Dirac sea continuum $(\omega')$ and the limits of the   Fermi sea continuum $(\omega_{up}\,, \omega_{low})$. b) The same for neutral kaons and antikaons. c)  Masses of $\pi^-\, \mbox{and}\, \pi^+$. The low energy mode, $\pi_s^+$ (lower curve) does not exist in the region of applicability of the model $(\rho\geq \rho_c)$.}
\end{center}
\end{figure}
%%%%%%%%%%%%%%%%%%%%%%%%%%%%%%%%%%%%

%%%%%%%%%%%%%%%%%%%%%%%%%%%%%%%%%%%%
\begin{figure}
\begin{center}
  \begin{tabular}{cc}
    \hspace*{-0.5cm}\epsfig{file=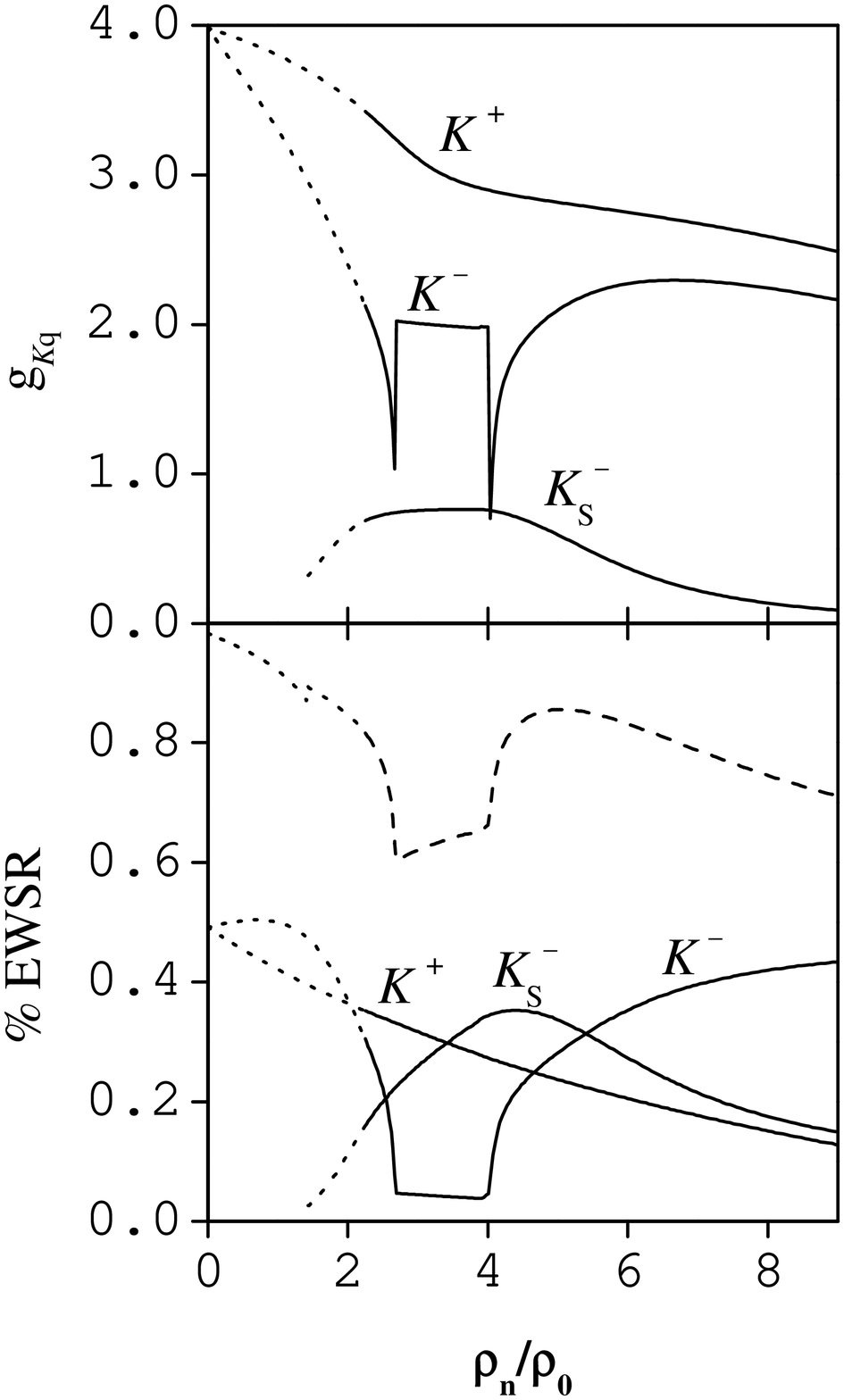,width=8.5cm,height=12cm} &
    \epsfig{file=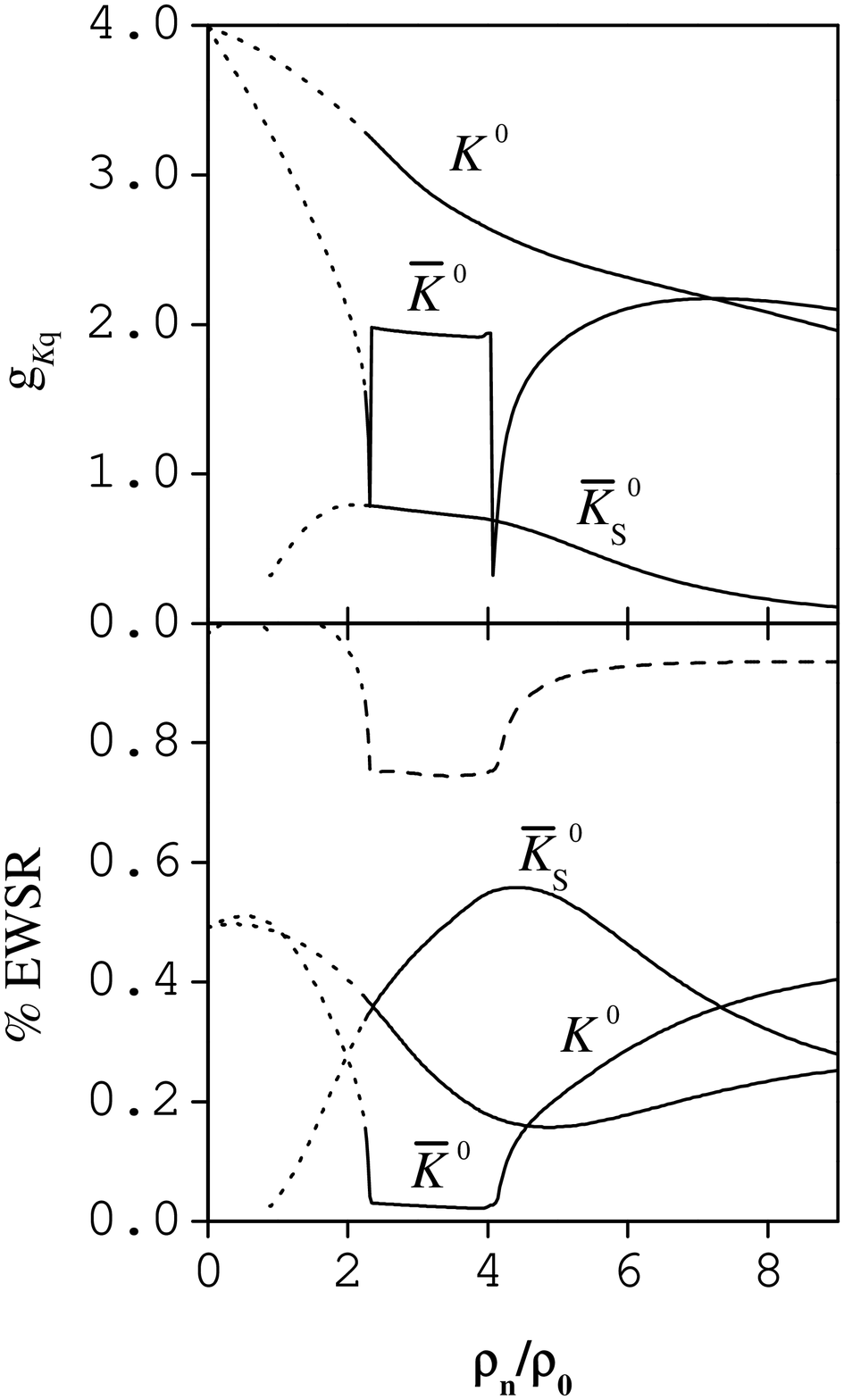,width=8.5cm,height=12cm} \\
  \end{tabular}
\caption{Quark - kaon coupling constant and strength (EWSR) of the modes; for $K^+$, $K^-$ (left) and $K^0$, $\bar K^0$ (right). The discontinuities in the strength of $K^-$ and $K^0$ occur in the range of densities where these modes are in the continuum ($2.25 - 4\rho_0$).}
\end{center} 
\end{figure}
%%%%%%%%%%%%%%%%%%%%%%%%%%%%%%%%%%%%

\end{document}